\documentclass{iaus}
\usepackage{graphicx}

\title[Pushing the precision of ground-based photometry] 
{Pushing the precision limit of ground-based eclipse photometry}

\author[M. Gillon et al.]   
{M. Gillon$^1$, D.R. Anderson$^2$, B.-O. Demory$^1$, D.M. Wilson$^2$, C. Hellier$^2$,  D. Queloz$^1$, C. Waelkens$^3$}

\affiliation{$^1$Observatoire de l'Universit\'e de Gen\` eve, Switzerland
\\ email: {\tt michael.gillon@obs.unige.ch} \\[\affilskip]
$^2$Astrophysics Group, Keele University, UK\\
$^3$Instituut voor Sterrenkunde, KU Leuven, Belgium}

\pubyear{2008}
\volume{253}  
\pagerange{1--7}
\jname{Transiting Planets}
\editors{D. Queloz \& D. Sasselov, eds.}
\begin{document}

\maketitle

\begin{abstract}
Until recently, it was considered by many that ground-based photometry could not 
reach the high cadence sub-mmag regime because of the presence 
of the atmosphere. Indeed, high frequency atmospheric noises 
(mainly scintillation) limit the precision that high SNR photometry 
can reach within small time bins. If one is ready to damage the 
sampling of his photometric time-series, binning the data (or using longer
exposures) allows to get better errors, but the obtained precision
will be finally limited by low frequency noises. To observe several times
the same planetary eclipse and to fold the photometry with the orbital
period is thus generally considered as the only option to get very
well sampled and precise eclipse light curve from the ground. 
Nevertheless, we show here that reaching the sub-mmag sub-min regime for
one eclipse is possible with a ground-based instrument. This has 
important implications  for transiting
planets characterization, secondary eclipses measurement and 
small planets detection from the ground.
\keywords{eclipses, techniques: photometric, planetary systems}
\end{abstract}

\firstsection 
\section{Introduction}

\begin{figure}[t]
\begin{center}
 \includegraphics[width=4.4in]{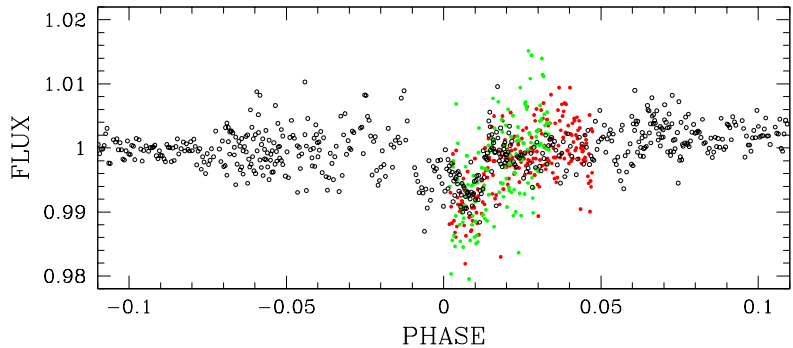} 
 \caption{Phase-folded GJ~436 OFXB photometry ($black$) from  \cite[Gillon et al. (2007c)]{Gillon07c}. }
   \label{fig1}
\end{center}
\end{figure}

Last year, the first transit of a `hot Neptune' was detected \cite[(Gillon et al. 2007c)]{Gillon07c}. This detection was not obtained with an expensive space instrument but with a commercial CCD camera mounted on a 60cm telescope located in Swiss mountains and mostly devoted to outreach activities. Most of the data were obtained in non-optimal transparency conditions (see Fig.\,\ref{fig1}) and are thus far to represent the best photometric quality that can be obtained with commercial equipment. Indeed, some amateur astronomers have demonstrated that they can obtain mmag transit photometry\footnote{see Bruce Gary's Amateur Exoplanet Archive {\tt http://brucegary.net/AXA/x.htm}} and they play an important role in the detection and characterization of transiting planets in the context of the {\tt TransitSearch.org} network \cite[(Barbieri et al. 2007)]{Barbieri07}  and the XO transit survey \cite[(Mc Cullough et al. 2006)]{xo2006}.

Detecting an eclipse shallower than 1\% is thus now possible with commercial equipment, but many efforts are undertaken to allow {\it the detection of eclipses shallower than 1 mmag} with professional ground-based instruments. It is indeed highly desirable to push the precision limit of ground-based photometry towards the sub-mmag regime. While we are presently able to detect and characterize from the ground gazeous giant planets transiting solar-type stars and Neptune-size planets transiting red dwarfs, what we should find below the mmag limit looks very exciting: transits of hot Neptunes around solar-type stars and of Super-Earths around M-dwarfs, secondary eclipse measurements in the visible and near-IR that would nicely complement the $Spitzer$ measurements, very accurate timing measurements allowing to detect Earth-mass planets via the Transit Timing Variations (TTV) method, and much more.

With a depth of 3 mmag, the shallowest eclipse detected so far from the ground is the transit of the core-dominated Saturn-mass planet HD\,149026b \cite[(Sato et al. 2005)]{Sato05}. The aim of this contribution can be summarized by a simple question: can we detect eclipses ten times smaller from the ground? 

\section{Observational and reduction strategy}

To reach the sub-mmag regime, several `rules of thumb' for high precision differential CCD photometry have to be known. Here is a summary:\begin{itemize}
\item {\it Pre-reduction}: the basic calibration of the raw images (flat-fielding, bias and dark subtraction, linearity and cosmetic correction) is of course important. A major issue is the quality of the flat-field, especially if the star images walk across the CCD during the run (dithering in the near-IR, poor guiding) and/or if the PSF is not spread over a huge number of pixels.
\item {\it PSF size and stability, reduction method}: for isolated stars much brigther than the background, the shape of the PSF and its stability across the field is not a big concern and using aperture photometry gives generally nice results. It is then better to defocus the telescope in order to minimize the impact of the different sensitivity of each pixel. In case of crowded fields and/or fainter stars, more sophisticated reduction methods based on PSF modeling are needed to get the best of the data, and the optical quality of the telescope and a good focusing become important. 
\item {\it Guiding}: keeping the star images on the same pixels is of course highly desirable to minimize the effect of the inter-pixel variability. In case of a highly structured and variable background (in the near-IR), it can nevertheless be preferable to dither frequently the telescope to build accurate sky maps to subtract to the images. 
\item {\it Choice of the pointing}: simply putting the target in the center of the field of view is rarely the best choice. The quality of the reference stars is a key factor in differential photometry, and it is worth spending time choosing the optimal pointing in order to get the best reference flux, i.e. as many non-variable stars of similar brightness and spectral type than the target as possible. 
\item {\it Good knowlegde of the noise budget}: to optimize the duty-cycle of the observations, the impact of the different sources of noise has to be properly estimated: photon noise of the target and the reference stars, scintillation, read-out noise and background noise (see e.g. \cite[Gilliland et al. 1993]{Gilliland93}). The noise contribution that is the most difficult to estimate $a$ $priori$ is the one due to correlated noises, whatever their origin (atmospheric, astronomical, instrumental). 
\end{itemize}

\section{Photometric quality indicators}

A term  such as `mmag-photometry' is misleading. With enough photons and a large enough time bin, it is easy to get a theoretical error that is close to 1 mmag, but this does not tell much about the quality of the resulting light curve. Three points have to be taken into account to estimate the quality of photometric data: \begin{itemize}
\item {\it Sampling (dT)}: this is a crucial point if one wants to determine very accurately the shape of an eclipse light curve (to, e.g., constraint thoroughly the impact parameter). An excellent sampling is also important to obtain very precise timing measurements (to, e.g., constraint the presence of another body in the system). Fast read-out camera are now largely widespread, and excellent duty cycle can be obtained with many instruments. In the near-IR,  the amplitude and variability of the background is a big concern for most targets. The resulting noise can be rather well corrected but for time bins equal or larger to the time scale of the background variability, generally ranging from some minutes to a few dozens of minutes. 
\item {\it Error per point ($\sigma$)}: it is obviously desirable to get a SNR as large as possible per measurement. For small aperture and/or time bin, scintillation can dominate the noise budget and it has to be taken into account to optimize the choice of the instrument and the strategy to use.
\item {\it Correlated noise ($\sigma_r$)}:  while the presence of low-frequency noises (due for instance to seeing variations or  an imperfect tracking) in any light curve was known since the prehistory of photometry, its impact on the final photometric quality has been often underestimated. This `red colored noise'  \cite[(Kruszewski \& Semeniuk 2003)]{Kru03} is nevertheless the actual limitation for high SNR photometric measurements  \cite[(Pont et al. 2006)]{Pont06} . The amplitude $\sigma_r$ of this `red noise' can be estimated from the residuals of the light curve itself \cite[(Gillon et al.  2006)]{Gillon06}, using:
\begin{equation}\label{eq:a}
\sigma_r =  \bigg(\frac{N\sigma_N^2 - \sigma^2}{N - 1}\bigg)^{1/2}\textrm{,}
\end{equation}

\noindent
where $\sigma$ is the $rms$ in the residuals and $\sigma_N$ is the standard deviation after binning these residuals into groups of $N$ points corresponding to a bin duration similar to the timescale of interest for an eclipse, the one of the ingress/egress.

\end{itemize}

\begin{figure}[t]
\begin{center}
 \includegraphics[width=5.5in]{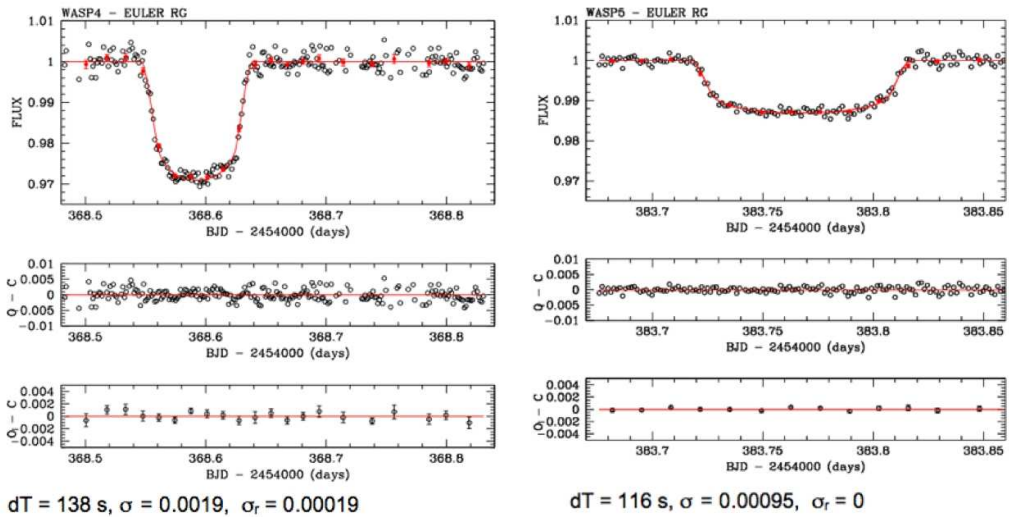} 
 \caption{Transit light curve obtained with the Euler Swiss telescope for WASP-4   \cite[(Wilson et al.  2008)]{Wilson08} and WASP-5 \cite[(Anderson et al.  2008)]{Anderson08}.}
 \label{fig2}
\end{center}
\end{figure}

Figure\,\ref{fig2} shows two transit light curves obtained with the Euler Swiss telescope (La Silla, Chile) during the characterization of the planets WASP-4  \cite[(Wilson et al.  2008)]{Wilson08} and WASP-5 \cite[(Anderson et al.  2008)]{Anderson08}. The corresponding values for $dT$, $\sigma$ and $\sigma_r$ are mentioned below each curve. These curves are representative of the photometric precision achieved routinely by several groups. We notice that using a good instrument and the rules of thumb presented in Section 2, a photometry nearly or even totally free of covariant noise can be obtained from a good astronomical site like La Silla.

\section{Towards the sub-mmag regime...}

To push farer the potential of ground-based eclipse photometry, three paths have been explored recently. All three seem very promising. \\

\begin{enumerate}

\item {\it Composite light curves} \\
One can take advantage of the periodic behavior of the eclipses to achieve very high photometric quality by combining multiple observations of the same eclipse. For the same sampling, both $\sigma$ and $\sigma_r$ should decrease as the square root of the number of observed eclipses (assuming that the covariant structures are not correlated for different eclipses). With a large enough number of observations, existing ground-based instruments are able to achieve a photometric quality that compares very well to what is obtained by space-based instruments. Figure\,\ref{fig3} shows for example the light curve resulting of the combination of 4 individual transits of GJ 436b that we observed with the Mercator Belgian Telescope located at La Palma. These transits were observed in the VG filter, the aim of these observations being to obtain an independent determination of the system parameters and to constraint the presence of another planet via the monitoring of the transit timings of the `hot Neptune' (Gillon et al. in prep.). With $dT$ = 47s, $\sigma$ = 810 ppm and not detectable covariant noise, this composite curve compares well with the individual transit light curve obtained at 8 $\mu$m with $Spitzer$ \cite[(Gillon et al.  2007b)]{Gillon07b} as shown on Fig.\,\ref{fig3}. MEROPE, the camera of Mercator, has a quite large read-out time of 60s, leading to a rather poor duty cycle for these observations, and we notice that the sampling of this composite curve would be significantly better with a state-of-the-art fast read-out camera. 

The `composite curve' approach relies on two basic assumptions: the perfect periodicity of the eclipse and the immutability of its shape. While these two assumptions are reasonable in most of the cases, It can be desirable to reach a very high precision for an individual eclipse, for instance in the case of an evolution of the orbital elements, TTVs or the presence of spots on the surface of the star.\\

\begin{figure}[t]
\begin{center}
 \includegraphics[width=4.7in]{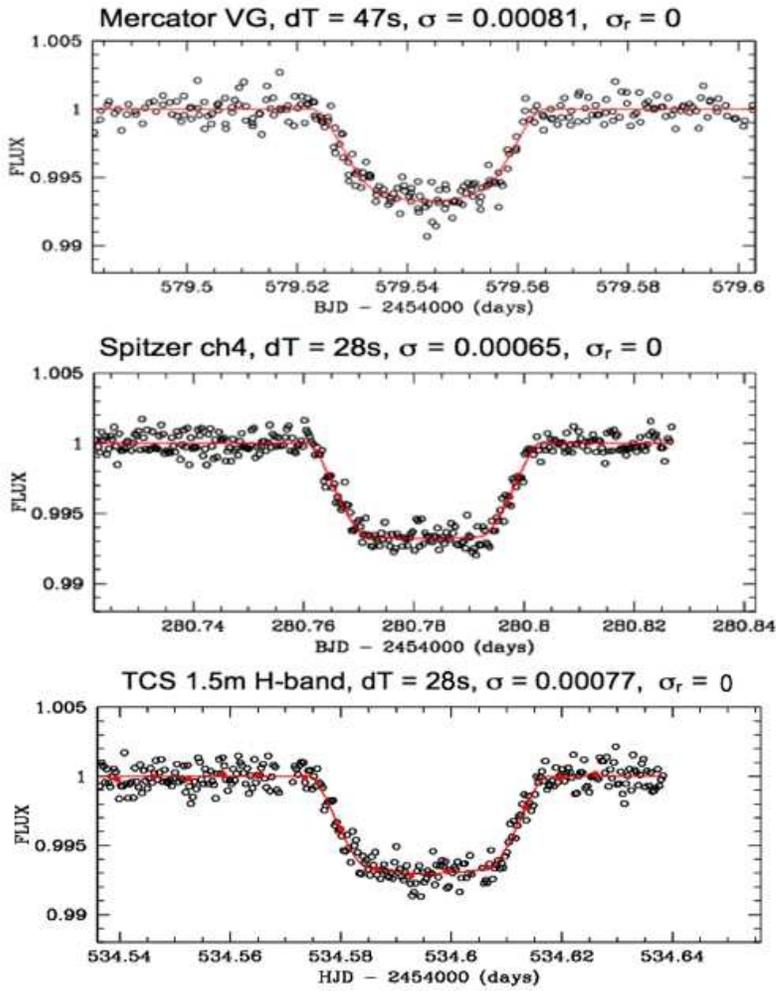} 
 \caption{Transit light curve obtained for GJ 436 with the Mercator Belgian telescope in the V-band ($Top$, composite curve, Gillon et al. in prep.), with $Spitzer$ at 8$\mu$m ($Middle$, Gillon et al. 2007b) and with TCS in the H-band ($Bottom$, Alonso et al. 2008).}
 \label{fig3}
\end{center}
\end{figure}

\item{\it Near-IR absolute photometry}

As outlined in Section 3, the amplitude and variability of the background is a major problem for high-precision highly-sampled ground-based near-IR photometry. But this problem vanishes if the brightness of the star is still much larger than the one of the background. Very recently, \cite[Alonso  et al.  (2008)]{Alonso08} observed a transit of GJ 436b in the H-band with the TCS telescope and its CAIN-II near-IR detector. As the red dwarf GJ 436 is very bright in the H-band (H = 6.3), no dithering pattern was used and the images were severely defocused, i.e. a strategy similar to what would be done in the visible was used. The baseline was corrected by a parabolic fit to the parts outside the transit. Figure\,\ref{fig3} compares the obtained light curve (binned to $\sim$ 28s) with the $Spitzer$ and the Mercator ones. The three curves have comparable values for $\sigma$ while they do not show any significant covariant noise.
 
The most surprising point here is that no differential photometry was used to reach this photometric quality.  \cite[Alonso  et al.  (2008)]{Alonso08}  explain this by the much smoother behavior of the transparency variations in the H-band compared to the visible. It is very desirable to confirm this claim by obtaining more high-quality eclipse light curves. Unfortunately, this method is limited to stars that are very bright in the near-IR, and only a few or them are known to harbor a transiting planet (e.g. HD\,189733, HD\,209458). \\

\item{\it Large telescopes for bright targets}

To reach a very precise and very well sampled photometry, the most obvious solution is to use the largest possible aperture. Photon noise and scintillation are not a concern even for small time bins when a rather bright transiting system is monitored with a large aperture telescope like the VLT. The high standard quality of such a telescope should allow one to avoid any instrumental systematic. We tested this simple approach by observing a transit of the planet WASP-4 with the VLT. We choose to observe in the $z$-filter to minimize the impact of the stellar limb-darkening on the deduced parameters. We observed this transit with the FORS2 camera that provides an excellent response in the red with a very low level of fringing. A very large defocus was used to obtain a good duty cycle and to minimize the influence of flat-fielding errors: the mean FHWM was 50 pixels = 12.5''. We outline that using such a large defocus is not at all a standard observational mode on the VLT. The defocus was tuned several times to adapt it to atmospheric transparency variations due to the increase of airmass. 

The first hour of data obtained at very low airmass suffers from a severe correlated noise. We  could not identify firmly the cause of the encountered problem. At this stage, we suspect that the problem is possibly linked to (1) an illumination problem of the VLT that should be stronger for low airmass observations, and (2)  the large defocus we used. Indeed, high-accuracy transit photometry has already been obtained with the FORS cameras (e.g.  \cite[Gillon et al. 2007a]{Gillon07a}), and the systematic presented here was not detected in these former data.  Fortunately, the transit occured in the second part of the run for which the effect seems to  be absent. We thus decided to reject the first part of data. The resulting transit light curve is shown in Fig. \,\ref{fig4}. The $rms$ of the first out-of-transit part is 420 ppm. This  value is very close to the theoretical error per point obtained from the photon noise of the target and the reference stars, the read-out noise and the scintillation noise: 400 ppm. For the second out-of-transit part, the measured $rms$ is 740 ppm while the median theoretical error is 510 ppm. This largest discrepancy between both values come probably from the amplification of the effect of any transparency inhomogeneity across the field at high airmass. Indeed, the airmass ranges from 1.45 to 1.95 in the second out-of-transit part. The analysis of the residuals lead to excellent photometric quality indicators: $dT$ = 54s , $\sigma$ = 550 ppm and $\sigma_r$ = 140 ppm.  

This `big telescope' approach has two major limitations. First of all, stars too bright would saturate the detector within very short times, even with large defocus, leading to poor duty cycle. Considering only this point , we estimate the optimum magnitude for the VLT/FORS2 instrument to be around V $\sim$ 11.5. Below, most of the observational time will be spent reading the detector. This problem could be solved with the use of a very short read-out time or frame-transfer CCD. The second limitation comes from the size of the field of view. For a magnitude below V $\sim$ 12, most of the targets would lack good reference stars in the 6.8' $\times$ 6.8' field of view of the FORS2 camera. For the VLT, a possible solution would be to observe the target with one telescope and a close-by reference star with another telescope, but such a strategy could be judged as very expensive, and only one reference star would not be enough to guarantee a very low level of correlated noise. 

\begin{figure}[t]
\begin{center}
 \includegraphics[width=3.4in]{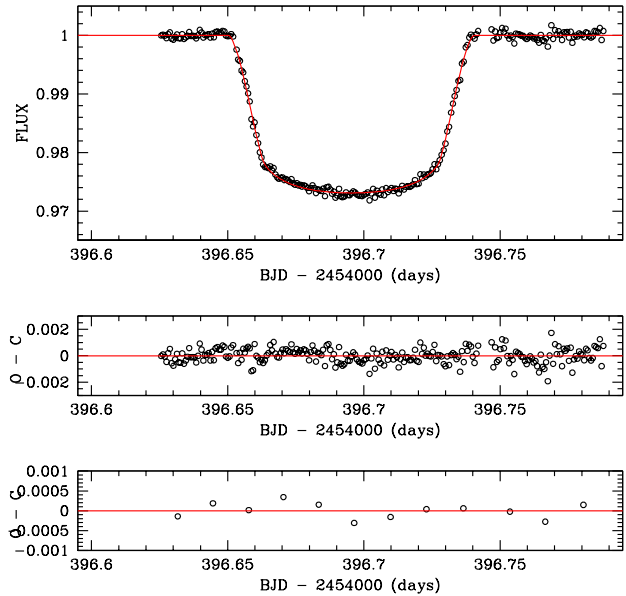} 
 \caption{$Top$: VLT/FORS2 $z$-band transit photometry for WASP-4. The best-fit transit curve is superimposed in red. $Middle$: residuals of the fit ($rms$ = 550 ppm). $Bottom$: residuals of the fit after binning per 20 points ($rms$ =  190 ppm).}
\label{fig4}
\end{center}
\end{figure}

\end{enumerate}

\section{Conclusion}

While the `near-IR' and `big telescope' approaches allow to get  high-precision highly-sampled eclipse photometry for one event observed from the ground but are limited to specific cases, the `composite light curve' approach has a much broader applicability. We are entering a new era of ground-based eclipse photometry, and we bravely predict that the first detection of a sub-mmag eclipse from the ground will be announced in the next future.

\end{document}